\documentclass{elsart}
\usepackage{graphicx,amssymb}
\begin{document}

\begin{frontmatter}

\title{ Multifractality and nonextensivity at the edge of chaos of unimodal maps} 
\author{E. Mayoral and A. Robledo}

\address{Instituto de F\'{i}sica,\\
Universidad Nacional Aut\'{o}noma de M\'{e}xico,\\
Apartado Postal 20-364, M\'{e}xico 01000 D.F., Mexico.\\ \small {E-mail addresses: estela@eros.pquim.unam.mx, robledo@fisica.unam.mx }}

\date{December 10, 2003}

\begin{abstract}
We examine both the dynamical and the multifractal properties at the chaos
threshold of logistic maps with general nonlinearity $z>1$. First we
determine analytically the sensitivity to initial conditions $\xi _{t}$.
Then we consider a renormalization group (RG) operation on the partition
function $Z$ of the multifractal attractor that eliminates one half of the
multifractal points each time it is applied. Invariance of $Z$ fixes a
length-scale transformation factor $2^{-\eta }$ in terms of the generalized
dimensions $D_{\beta }$. There exists a gap $\Delta \eta $ in the values of $%
\eta $ equal to $\lambda _{q}=1/(1-q)=D_{\infty }^{-1}-D_{-\infty }^{-1}$
where $\lambda _{q}$ is the $q$-generalized Lyapunov exponent and $q$ is the
nonextensive entropic index. We provide an interpretation for this
relationship - previously derived by Lyra and Tsallis - between dynamical
and geometrical properties.

\end{abstract}


\end{frontmatter}

\section{Introduction}

The thermodynamic framework for the characterization of multifractals \cite
{beck1} has proved to be a useful and insightful tool for the analysis of
many complex system problems in which this geometrical feature plays a
prominent role in establishing physical behavior. This has clearly been the
case of dynamical systems, turbulence, growth models, sand pile models, etc. 
\cite{multifractal1}. As it has recently been pointed out, multifractality
is also of relevance to critical \cite{antoniou1} and glassy dynamics \cite
{robledo1}. In the former case the geometry of critical fluctuations is
found to be multifractal and this has a hold on their dynamical properties 
\cite{antoniou1}, whereas in the latter case slow dynamics develops as
phase-space accessibility is reduced and is finally confined to a
multifractal subspace \cite{robledo1}. Concurrently, in the study of the
same types of systems the nonextensive generalization of the Boltzmann Gibbs
(BG) statistical mechanics \cite{tsallis0}, \cite{tsallis1} has recently
raised much interest and provoked considerable debate \cite{science1} as to
whether there is firm evidence for its applicability in circumstances where
a system is out of the range of validity of the canonical BG theory. Slow
dynamics is plausibly related to hindrance of movement in phase space that
leads to ergodicity breakdown, and this in turn has been linked to the
failure of BG statistics \cite{robledo1}. Here we briefly examine the
relationship between multifractal properties and nonextensive dynamical
evolution at the edge of chaos, a nonergodic state, in unimodal maps, as
part of a wider analysis \cite{robledo2}, that might contribute to reveal
the physical source of the nonextensive statistics.

\section{Nonextensivity of the Feigenbaum attractor}

An important class of multifractals are the strange attractors that are
generated by iteration of nonlinear maps. Amongst these, a prototypical
example is the Feigenbaum attractor, also known as the edge of chaos,
generated by the logistic map , $f_{\mu ,2}(x)=1-\mu x^{2}$,$\;-1\leq x\leq
1 $, when the control parameter $\mu $ takes the value $\mu _{\infty
}=1.40115...$ \cite{beck1}. Recently, the predictions of the nonextensive
theory have been rigorously proved \cite{baldovin1} \cite{baldovin2} for
this state that marks the transition from periodic to chaotic motion and
that is characterized by the vanishing of the ordinary Lyapunov exponent $%
\lambda _{1}=0$. At the edge of chaos (and also at infinitely many other
critical states of the same map such as the pitchfork and tangent
bifurcations where $\lambda _{1}=0$ \cite{baldovin3}) the sensitivity to
initial conditions $\xi _{t}\equiv \left| dx_{t}/dx_{0}\right| $ (where $%
x_{t}$ is the orbit position at time $t$ given the initial position $x_{0}$
at time $t=0$) generally acquires power law instead of exponential time
dependence. The nonextensive formalism suggests \cite{tsallis2} that $\xi
_{t}$ when $\lambda _{1}=0$ obeys the $q$-exponential expression, 
\begin{equation}
\xi _{t}=\exp _{q}(\lambda _{q}t)\equiv [1-(q-1)\lambda _{q}t]^{-1/(q-1)},
\label{sensitivity}
\end{equation}
containing the entropic index $q$ and the $q$-generalized Lyapunov
coefficient $\lambda _{q}$. The BG exponential form $\xi _{t}=\exp (\lambda
_{1}t)$ is recovered when $q\rightarrow 1$. Further, it was conjectured \cite
{tsallis2} that the ordinary Pesin identity $\lambda _{1}=K_{1}$, where $%
K_{1}$ is the Kolmogorov-Sinai (KS) rate of entropy increase \cite{note1} is
replaced by $\lambda _{q}=K_{q}$, where $K_{q}t=S_{q}(t)-S_{q}(0)$, $t$
large, and where 
\begin{equation}
S_{q}\equiv \sum_{i}p_{i}\ln _{q}\left( \frac{1}{p_{i}}\right) =\frac{%
1-\sum_{i}^{W}p_{i}^{q}}{q-1}  \label{entropy}
\end{equation}
is the Tsallis entropy ($\ln _{q}y\equiv (y^{1-q}-1)/(1-q)$ is the inverse
of $\exp _{q}(y)$). In the limit $q\rightarrow 1$ $K_{q}$ becomes $%
K_{1}\equiv t^{-1}[S_{1}(t)-S_{1}(0)]$ where $S_{1}(t)=-%
\sum_{i=1}^{W}p_{i}(t)\ln p_{i}(t)$. In Eq. (\ref{entropy}) $p_{i}(t)$ is
the probability distribution obtained from the relative frequencies with
which the positions of an ensemble of trajectories occur within cells $%
i=1,...,W$ at iteration time $t$. Based on the renormalization group (RG)
transformation ${\bf R}f(x)\equiv \alpha f(f(x/\alpha ))$ where $\alpha
\simeq 2.5029$ is the Feigenbaum's universal constant that measures the
power-law period-doubling spreading of iterate positions, a rigorous
analytical confirmation of Eq. (\ref{sensitivity}) and also of $\lambda
_{q}=K_{q}$ have been obtained \cite{baldovin1} \cite{baldovin2}.
Specifically, $\lambda _{q}$ and $q$ are simply given by $\lambda _{q}=\ln
\alpha /\ln 2\simeq 1.3236$ and $q=1-\ln 2/\ln \alpha \simeq 0.2445$.

On the other hand, a remarkable link between the well-known \cite{beck1}
geometrical description of multifractals and the nonextensive statistics has
been derived and corroborated numerically not only for the Feigenbaum
attractor but for the edge of chaos of the generalization of the logistic
map to nonlinearity $z>1$, $f_{\mu ,z}(x)=1-\mu \left| x\right| ^{z}$,$%
\;-1\leq x\leq 1$ \cite{tsallis3}. This consists of the expression 
\begin{equation}
\frac{1}{1-q}=\frac{1}{D_{\infty }}-\frac{1}{D_{-\infty }},  \label{tsallis}
\end{equation}
where $D_{\infty }$ and $D_{-\infty }$ are the extreme values of the
generalized dimension spectrum $D_{\beta }$, and correspond to the smallest
and largest length scales of the multifractal set \cite{beck1}. For the $z$%
-logistic map one has $D_{\infty }=\ln 2/z\ln \alpha $ and $D_{-\infty }=\ln
2/\ln \alpha $ (notice that $\alpha $ depends on $z$) \cite{beck1}. In view
of the clear association of the nonextensive statistics with the properties
at the edge of chaos of the logistic map we have pursued the study of this
critical state by means of the standard thermodynamic tools that
characterize multifractals. Here we advance some of the results we have
found \cite{robledo2}.

\section{Sensitivity to initial conditions}

We determine the index $q$ and the $q$-Lyapunov coefficient $\lambda _{q}$
at the edge of chaos of the $z$-logistic maps through direct evaluation of
the sensitivity to initial conditions $\xi _{t}$. By making use of the
properties \cite{vanderweele1} $g(0)=(-\alpha )^{n}g^{(2^{n})}(0)=1$, $%
g(1)=(-\alpha )^{n}g^{(2^{n})}(\alpha ^{-n})=-\alpha ^{-1}$ and 
\begin{equation}
g^{\prime }(1)=(-1)^{n}g^{\prime }(\alpha ^{-n})g^{\prime }(g(\alpha
^{-n}))...g^{\prime }(g^{(2^{n}-1)}(\alpha ^{-n}))=-\alpha ^{z-1},
\end{equation}
of the fixed-point map, $g(x)=\lim_{n\rightarrow \infty }(-\alpha
)^{n}f_{\mu ,z}^{(2^{n})}(-x/\alpha ^{n})$, into the definition of $\xi _{t}$%
, 
\begin{equation}
\xi _{t}\equiv \left| \frac{dx_{t}}{dx_{0}}\right| =\left| \frac{%
dg^{(2^{n}-1)}(x)}{dx}\right| _{x=x_{0}},
\end{equation}
with the choice $x_{0}=0$, and for iteration times of the form $t=2^{n}-1$,
we obtain $\xi _{t}=\alpha ^{(z-1)n}$. This result can be re-expressed as 
\begin{equation}
\xi _{t}=\exp _{q}(\lambda _{q}t)\equiv [1-(q-1)\lambda
_{q}t]^{-1/(q-1)},\quad t=2^{n}-1,
\end{equation}
with 
\begin{equation}
\lambda _{q}=\frac{1}{1-q}=(z-1)\frac{\ln \alpha }{\ln 2}.
\label{q-lyapunov}
\end{equation}

For details see Ref. \cite{robledo2}. Thus, without reference to the
multifractal properties of the attractor, we have corroborated the result in
Eq. (\ref{tsallis}) \cite{tsallis3} and obtained additionally the $q$%
-Lyapunov coefficient $\lambda _{q}=(1-q)^{-1}$. Alternatively, $\xi _{t}$
can be obtained from $\left| \Delta x_{t}\right| /\left| \Delta x_{0}\right| 
$ in the limit $\Delta x_{0}\rightarrow 0$ where $\Delta x_{t}$ is the
distance between two orbits at time $t$ when they were a distance $\Delta
x_{0}$ at $t=0$. Following the procedure described in Ref. \cite{baldovin1}
with the choices $\left| \Delta x_{0}\right| =\alpha ^{-j}-\alpha ^{-i}$, $%
i,j>0$ and $t=2^{n}-1$, $n=0,1,...$, we obtain $\left| \Delta x_{t}\right| =$
$\left| \Delta x_{0}\right| \alpha ^{(z-1)n}$ \cite{robledo2}. Next we study
a multifractal property of the attractor and make contact again with Eq. (%
\ref{tsallis}).

\begin{figure}
\includegraphics[width=6.5cm,height=6.8cm,angle=270]{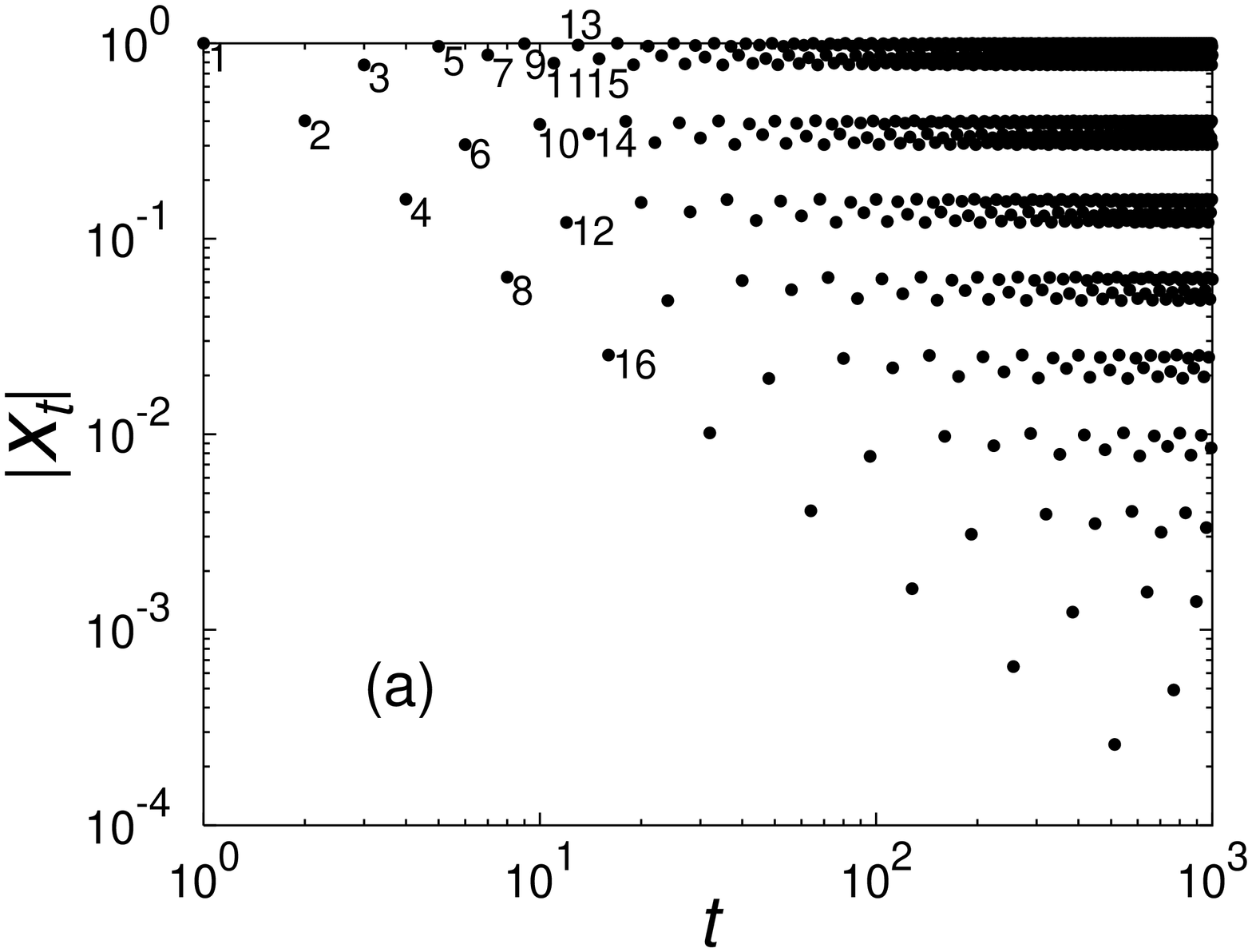}
\includegraphics[width=6.5cm,height=6.6cm,angle=270]{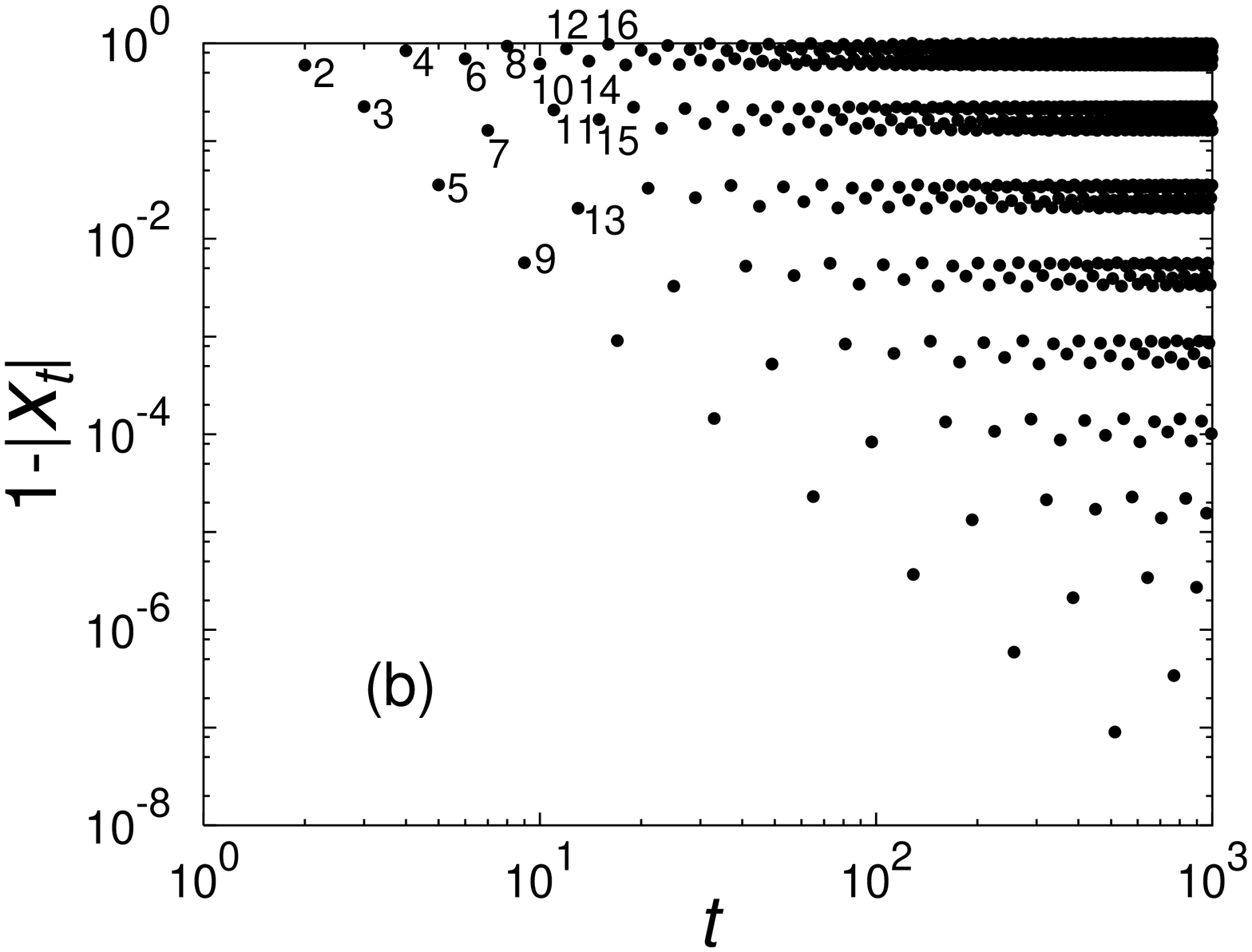}
\caption{a) Absolute values of $\left| x_{t}\right| ${\ vs }$t$%
{\ in logarithmic scales for the orbit with initial condition }$%
x_{0}=0${ \ at }$\mu _{\infty }${\ of the logistic map }$z=2$%
{. The labels indicate iteration time }$t${. b) Same as a)
with }$\left| x_{t}\right| ${\ replaced by }$1-\left| x_{t}\right| $%
{.}}
\label{fig1}
\end{figure}

\section{Scaling properties at the edge of chaos}

In Fig. 1a we have plotted (in logarithmic scales) the first few (absolute
values) of iterated positions $\left| x_{t}\right| $ of the orbit at $\mu
_{\infty }$ starting at $x_{0}=0$ where the labels indicate iteration time $%
t $. Notice the structure of horizontal bands, and that, in the top band lie
half of the attractor positions (odd times), in the second band a quarter of
the attractor positions, and so on. The top band is eliminated by functional
composition of the original map, that is by considering the orbit generated
by the map $g^{(2)}(0)$ instead of $g(0)$. Successive bands are eliminated
by considering the orbits of $g^{(2^{k})}(0),\ k=1,2...$. The top band
positions (odd times) can be reproduced approximately by the positions of
the band below it (times of the form $t=2+4n$, $n=0,1,2...$) by
multiplication of a factor equal to $\alpha $, e.g. $\left| x_{1}\right|
\simeq \alpha \left| x_{2}\right| $. Likewise, the positions of the second
band are reproduced by the positions of the third band under multiplication
by $\alpha $, e.g. $\left| x_{2}\right| \simeq \alpha \left| x_{4}\right| $,
and so on. In Fig. 1b we show a logarithmic scale plot of $1-$ $\left|
x_{t}\right| $ that displays a band structure similar to that in Fig. 1a, in
the top band lie again half of the attractor positions (even times) and
below the other half (odd times) is distributed in the subsequent bands.
This time the positions in one band are reproduced approximately from the
positions of the band lying below it by multiplication of a factor $\alpha
^{z}$, e.g. $1-\left| x_{3}\right| \simeq \alpha ^{z}(1-\left| x_{5}\right|
) $. The scaling properties amongst iterate positions merely follow from
repeated composition of the map (for more details see Ref. \cite{robledo2})
and suggest the construction of an RG transformation described below. For
this purpose it is convenient to write the two scale factors $\alpha $ and $%
\alpha ^{z}$ as $2^{\eta _{-\infty }}$ and $2^{\eta _{\infty }}$,
respectively, where $\eta _{-\infty }=\ln \alpha /\ln 2$ and $\eta _{\infty
}=z\ln \alpha /\ln 2$.

\section{'Take away half' RG transformation}

Consider the positions of the orbit at $\mu _{\infty }$ starting at $x_{0}=0$
up to time $t=2^{N}-1$ and from them extract the set of lengths $%
l_{j}^{(0)},\ j=1...,2^{N}$, that join closest adjacent positions. Each
position is used only once so that the set $l_{j}^{(0)}$ covers the
attractor but not all the interval $-1\leq x\leq 1$. To each length $%
l_{j}^{(0)}$ assign the equiprobability $p_{j}^{(0)}=2^{-N}$. Following the
standard procedure \cite{beck1} we define the partition function 
\begin{equation}
Z_{N}^{(0)}(\beta ,\tau )=\sum_{j}[p_{j}^{(0)}]^{\beta }[l_{j}^{(0)}]^{\tau
},  \label{partition0}
\end{equation}
and determine the (finite size) generalized dimensions $D_{\beta ,N}\equiv
\tau /(\beta -1)$ by requiring $Z_{N}^{(0)}(\beta ,\tau )=1$, ($D_{\beta
}=\lim_{N\rightarrow \infty }D_{\beta ,N}$). In Fig. 2 we show results for $%
D_{\beta ,N}$ for several values of $z$.

\begin{figure}
\begin{center}
\includegraphics[width=6cm
,angle=270]{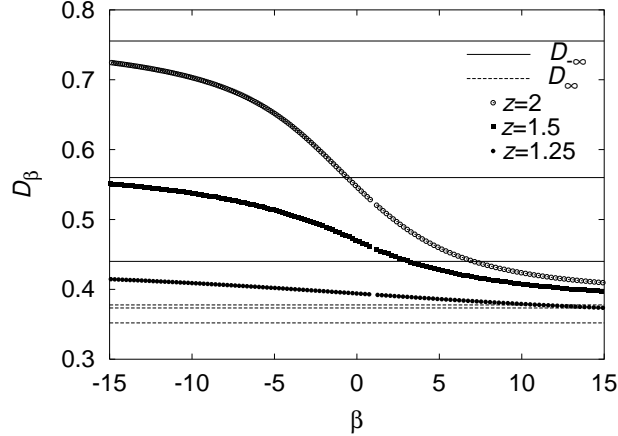}
\end{center}
\caption{{\small The generalized dimensions }$D_{\beta ,N}${\small \ for }$%
z=1.25${\small \ (}$\alpha (1.25)=4.8323,\ N=13${\small ), }$z=1.5${\small \
(}$\alpha (1.5)=3.4479,\ N=16${\small ) and }$z=2${\small \ (}$\alpha
(2)=2.5029,\ N=14${\small ).}
}
\label{fig2}
\end{figure}

We introduce now repeated length scale and restoring probability
transformations 
\begin{equation}
l_{j}^{(k)}=2^{-k\eta }l_{j}^{(0)}\quad \textrm{and}\quad
p_{j}^{(k)}=2^{-k}p_{j}^{(0)},\quad k=1...,2^{N},  \label{transformations}
\end{equation}
on the partition function $Z_{N}^{(0)}(\beta ,\tau )$. That is, to transform 
$Z_{N}^{(0)}(\beta ,\tau )$ into $Z_{N}^{(1)}(\beta ,\tau )$ replace $%
l_{j}^{(0)}$ and $p_{j}^{(0)}$ in Eq. (\ref{partition0}) by $%
l_{j}^{(1)}=2^{-\eta }l_{j}^{(0)}$ and $p_{j}^{(1)}=2^{-1}p_{j}^{(0)}$,
respectively, and proceed similarly any number $k$ of times. In the limit $%
N\rightarrow \infty $ the transformation eliminates one half of the
multifractal points each time it is applied. By requiring $Z_{N}^{(k)}(\beta
,\tau )=1$ we obtain, again when $N\rightarrow \infty $, the following
result for the rescaling exponent $\eta $ of the interval lengths 
\begin{equation}
\eta =\frac{\beta }{\tau }=\frac{\beta }{\beta -1}\frac{1}{D_{\beta }}.
\label{eta}
\end{equation}

\begin{figure}
\begin{center}
\includegraphics[width=6cm,angle=270]{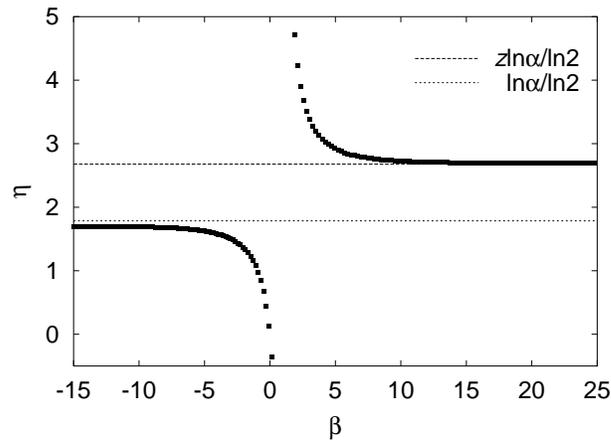}
\end{center}
\caption{
{\small The length rescaling index }$\eta ${\small \ in Eq. (\ref
{eta}) for }$z=1.5${\small \ (}$\alpha (1.5)=3.4479,\ N=16${\small ).}
}
\label{fig3}
\end{figure}

As shown in Fig. 3, $\eta $ approaches two limiting values, $\eta _{-\infty
}=D_{-\infty }^{-1}$ and $\eta _{\infty }=D_{\infty }^{-1}$ as $\beta
\rightarrow -\infty $ and $\beta \rightarrow \infty $, respectively, and
exhibits a singularity at $\beta =1$. Except for values of $\beta $ in the
vicinity of $\beta =1$ the exponent $\eta $ is always close to either $\eta
_{-\infty }$ or $\eta _{\infty }$ which correspond to the scaling factors $%
l_{j}^{(k)}=\alpha ^{-k}l_{j}^{(0)}$ and $l_{j}^{(k)}=\alpha
^{-zk}l_{j}^{(0)}$ described in the previous section. As also shown in Fig.
3, there appears a gap in the permissible values of $\eta $, 
\begin{equation}
\Delta \eta \equiv \eta _{\infty }-\eta _{-\infty }=\frac{1}{D_{\infty }}-%
\frac{1}{D_{-\infty }},  \label{etagap}
\end{equation}
that is noticeably identical to the dynamical result Eq. (\ref{q-lyapunov}),
i.e. $\lambda _{q}=(1-q)^{-1}=\Delta \eta $.

Repeated application of the RG transformation when the limit $N\rightarrow
\infty $ is taken before $k\rightarrow \infty $ leaves the generalized
dimensions invariant, i.e. $D_{\beta }=D_{\beta }^{(k)}$ for all $k$.
However when both $k,N\rightarrow \infty $ with $k=N$ the RG transformation
leads to a fixed point $D_{\beta }^{(k)}\rightarrow D_{\beta }^{(\infty )}$
that has the form of a step function, $D_{\beta }^{(\infty )}=\ln 2/2\ln
\alpha $ for $\beta <1$ and $D_{\beta }^{(\infty )}=\ln 2/2z\ln \alpha $ for 
$\beta >1$, with a divergence at $\beta =1$. See Fig. 4. \cite{robledo2}.

\begin{figure}
\begin{center}
\includegraphics[width=7cm,angle=270]{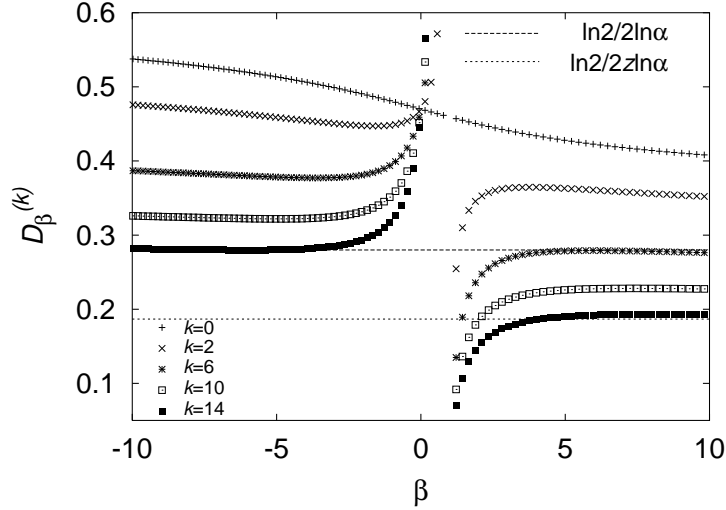}
\end{center}
\caption{{\small Successively RG transformed dimensions }$D_{\beta }^{(k)}$%
{\small \ for }$k/N=1${\small \ and }$z=1.5${\small .}
}
\label{fig4}
\end{figure}

\section{Summary and discussion}

Assisted by the known properties of the fixed-point map $g(x)$ we derived
the expression for the sensitivity to initial conditions $\xi _{t}$ for the $%
z$-logistic map at the edge of chaos. As a consequence we corroborated the
known expression for the entropic index $q$ in terms of $\alpha (z)$ \cite
{tsallis3} and obtained in addition the closely related expression for the $%
q $-Lyapunov coefficient $\lambda _{q}$ (known previously only for the
particular case $z=2$ \cite{baldovin1} \cite{baldovin2}). Our results are
rigorous. The expression obtained is precisely that first derived,
heuristically, by Lyra and Tsallis \cite{tsallis3}, Eq. (\ref{tsallis}),
that relates $q$ to the smallest $D_{\infty }$ and largest $D_{-\infty }$
generalized dimensions of a multifractal set. In order to probe this
connection between dynamic and geometric properties of the multifractal
attractor we introduced an RG approach that consists of sequentially
removing remaining halves $2^{-k},\ k=1,2,...$ of points of the multifractal
set while keeping the value of its partition function $Z_{N}^{(k)}(\beta
,\tau )$ fixed. We determined the length rescaling index $\eta $ as a
function of $\beta $ and observed that there is a gap $\Delta \eta $ of
permissible values in this quantity equal to $\lambda _{q}=1/(1-q)$, as the
edges of the gap correspond to $\eta _{-\infty }=D_{-\infty }^{-1}$ and $%
\eta _{\infty }=D_{\infty }^{-1}$. This result can perhaps be better
appreciated if we note that the orbit positions $\left| x_{t}\right| $
evolve according to two basic alternating movements observable at time
intervals of the form $2^{n}$, one drives the iterate towards $\left|
x\right| =1$ as $\alpha ^{-zn}=2^{-\eta _{\infty }\ n}$ and the other one
towards $x=0$ as $\alpha ^{-n}=2^{-\eta _{-\infty }\ n}$. See the main
diagonal subsequence positions in Figs. 1a and 1b. The net growth in the
distance between two orbits is the ratio of these two power laws, that is $%
\left| \Delta x_{t}\right| $ $=$ $\left| \Delta x_{0}\right| 2^{\Delta \eta
\ n}$ (see Section 3). Thus, Eq. (\ref{tsallis}) is obtained as the result
of orbits that successively separate and converge, respectively, in relation
to $D_{-\infty }$ and $D_{\infty }$. The power law for $\xi _{t}$ can only
persist if $D_{-\infty }\neq D_{\infty }$.

Acknowledgments. We were partially supported by CONACyT grant P40530-F
(Mexican agency).


\begin{thebibliography}{99}
\bibitem{beck1}  See, for example, C. Beck and F. Schlogl, {\it %
Thermodynamics of Chaotic Systems} (Cambridge University Press, UK, 1993).

\bibitem{multifractal1}  See, for example, M. Schroeder, {\it Fractals,
Chaos, Power Laws} (W.H.Freeman and Company, NY, 1991).

\bibitem{antoniou1}  N.G. Antoniou, Y.F. Contoyiannis and F.K. Diakonos,
Phys. Rev. E 62, 3125 (2000).

\bibitem{robledo1}  A. Robledo, cond-mat/0307285.

\bibitem{tsallis0}  C. Tsallis, J. Stat. Phys. 52, 479 (1988).

\bibitem{tsallis1}  For a recent review see, C. Tsallis, A. Rapisarda, V.
Latora and F. Baldovin, in {\it Dynamics and Thermodynamics of Systems with
Long-Range Interactions, }eds. S. Ruffo, E. Arimondo and M. Wilkens, Lecture
Notes in Physics 602, 140 (Springer, Berlin, 2002). See
http://tsallis.cat.cbpf.br/biblio.htm for full bibliography.

\bibitem{science1}  A. Cho, Science 297, 1269 (2002); and Letters to the
Editor, Science 300, 249 (2003).

\bibitem{robledo2}  E. Mayoral and A. Robledo, to be submitted for
publication.

\bibitem{baldovin1}  F. Baldovin and A. Robledo, Phys. Rev. E 66, 045104(R)
(2002).

\bibitem{baldovin2}  F. Baldovin and A. Robledo, cond-mat/0304410.

\bibitem{baldovin3}  F. Baldovin and A. Robledo, Europhys. Lett. 60, 518
(2002); A. Robledo, Physica D (2003, in press) and cond-mat/0202095.

\bibitem{tsallis2}  C. Tsallis, A.R. Plastino and W.-M. Zheng, Chaos,
Solitons and Fractals 8, 885 (1997).

\bibitem{note1}  Here we refer to a simple form of rate of entropy increment
that differs from the usual definition of the KS entropy (see Ref. \cite
{baldovin2}).

\bibitem{tsallis3}  M. Lyra and C. Tsallis, Phys. Rev. Lett. 80. 53 (1998).

\bibitem{vanderweele1}  J.P van der Weele, H.W. Capel and R. Kluiving,
Physica A 145, 425 (1987).\newpage 
\end{thebibliography}
\end{document}